\documentclass[journal=jacsat,manuscript=article]{achemso}
\usepackage[utf8]{inputenc}        
\DeclareUnicodeCharacter{2212}{-}   

\usepackage[english]{babel}
\usepackage{graphicx}
\usepackage{bm}
\usepackage{float}
\usepackage[percent]{overpic}
\usepackage{braket}
\usepackage{amsmath}
\usepackage{amssymb}
\usepackage{gensymb}
\usepackage[utf8]{inputenc} % set input encoding to utf8

\floatstyle{plaintop}
\restylefloat{table}

\usepackage{achemso}
\usepackage[version=4]{mhchem}
\usepackage[%
  colorlinks=true,
  urlcolor=black,
  linkcolor=black,
  citecolor=black
]{hyperref}
\usepackage{breqn}
\usepackage{multirow}
\usepackage[nottoc]{tocbibind}

\usepackage{makecell}        % for \makecell line-breaks
\usepackage[table]{xcolor}   % for row/col colours
\definecolor{headergrey}{HTML}{C0C0C0}
\definecolor{rowgrey}{HTML}{EFEFEF}

\usepackage{dcolumn}
\usepackage{placeins}
\usepackage{microtype}
\usepackage[font=small,labelfont=bf]{caption}
\usepackage[tableposition=top]{caption}
\usepackage{graphicx}
\usepackage[percent]{overpic}  % percentage
\usepackage{xcolor}
\usepackage{setspace} 
\usepackage{etoolbox}

\makeatletter
\renewcommand*{\acs@author@fnsymbol@symbol}[1]{
    \ifcase #1 *\or
    a\or
    b\or
    c\or
    d\or
    e\or
    f\or
    g\or
    h\or
    i\or
    j
    \fi
}
\renewcommand*\acs@contact@details{
    {\sffamily *\,E-mail: \acs@email@list }%
    \acs@number@list
}  
\patchcmd{\acs@address@list@auxii}
{\acs@author@fnsymbol{\acs@affil@marker@cnt}}
{\textsuperscript{\acs@author@fnsymbol{\acs@affil@marker@cnt}}}
{}{}
\patchcmd{\acs@address@list@auxii}
{{\acs@author@fnsymbol{\acs@affil@marker@cnt}\@nameuse{@altaffil@\@roman\@tempcnta}\par}}
{{\textsuperscript{\acs@author@fnsymbol{\acs@affil@marker@cnt}}\@nameuse{@altaffil@\@roman\@tempcnta}\par}}
{}{} 

\title{Anomalous double-layer restructuring in water-in-salt electrolytes at graphitic interfaces governs capacitance}

\author{Hannah O.\ Wood}
\email{hannah.wood-3@manchester.ac.uk}
\affiliation[UoM Chemistry]{Department of Chemistry, University of Manchester, Oxford Road, Manchester M13 9PL, United Kingdom}
\alsoaffiliation[UoM ChemEng]{Department of Chemical Engineering, University of Manchester, Oxford Road, Manchester M13 9PL, United Kingdom}

\author{Fulu Zhou}
\affiliation[UoM Chemistry]{Department of Chemistry, University of Manchester, Oxford Road, Manchester M13 9PL, United Kingdom}
\alsoaffiliation[UoM ChemEng]{Department of Chemical Engineering, University of Manchester, Oxford Road, Manchester M13 9PL, United Kingdom}

\author{Jan Do\v ckal}
\affiliation[UJEP Physics]{Department of Physics, Faculty of Science, Jan Evangelista Purkyn\v e University in \v Ust\'i nad Labem, Pasteurova 3544/1, 400 96 \v Ust\'i nad Labem, Czech Republic}

\author{Martin L\'isal}
\affiliation[ICPF CAS]{Research Group of Molecular and Mesoscopic Modelling, Institute of Chemical Process Fundamentals, Czech Academy of Sciences, Rozvojov\'a 135/1, 165 02 Prague, Czech Republic}
\alsoaffiliation[UJEP Physics]{Department of Physics, Faculty of Science, Jan Evangelista Purkyn\v e University in \v Ust\'i nad Labem, Pasteurova 3544/1, 400 96 \v Ust\'i nad Labem, Czech Republic}

\author{Filip Mou\v cka}
\affiliation[UJEP Physics]{Department of Physics, Faculty of Science, Jan Evangelista Purkyn\v e University in \v Ust\'i nad Labem, Pasteurova 3544/1, 400 96 \v Ust\'i nad Labem, Czech Republic}

\author{Sittipong Kaewmorakot}
\affiliation[UoM Chemistry]{Department of Chemistry, University of Manchester, Oxford Road, Manchester M13 9PL, United Kingdom}
\alsoaffiliation[Henry Royce]{Henry Royce Institute, University of Manchester, Oxford Road, Manchester M13 9PL, United Kingdom}

\author{Robert A.\ W.\ Dryfe}
\affiliation[UoM Chemistry]{Department of Chemistry, University of Manchester, Oxford Road, Manchester M13 9PL, United Kingdom}
\alsoaffiliation[Henry Royce]{Henry Royce Institute, University of Manchester, Oxford Road, Manchester M13 9PL, United Kingdom}

\author{Paola Carbone}
\email{paola.carbone@manchester.ac.uk}
\affiliation[UoM Chemistry]{Department of Chemistry, University of Manchester, Oxford Road, Manchester M13 9PL, United Kingdom}

\begin{document}
\begingroup
\makeatletter
\let\orig@fnsymbol\@fnsymbol
\makeatother

\begin{abstract} The structure and thickness of the electrical double layer (EDL) at carbon electrodes strongly influence electrochemical performance, yet remain poorly understood in super-concentrated aqueous electrolytes. Here we combine classical and quantum-mechanical molecular dynamics simulations to resolve the interfacial organisation of aqueous LiCl from dilute to water-in-salt (WiS) (1--$20~\mathrm{mol~kg^{-1}}$) concentrations at graphitic electrodes, and compare with electrochemical differential-capacitance measurements from which the potential of zero charge (PZC) is obtained. We uncover a concentration-driven restructuring of the EDL: below $6~\mathrm{mol~kg^{-1}}$, solvated Li$^+$ dominates the outer Helmholtz plane (OHP), but at higher concentrations co-adsorption of Cl$^-$ through solvent-separated ion pairs enforces a near 1:1 Li:Cl ratio at the interface. This transition expands the effective EDL thickness, redistributes the interfacial potential drop, and drives a decrease in the PZC, matching the trend inferred from differential-capacitance measurements on electrolyte-graphite interfaces. Capacitance calculations reveal that while both EDL and quantum contributions vary strongly with concentration, their opposing trends make the total capacitance appear nearly constant for pristine few-layer graphite; for electrodes with smaller quantum capacitance, however, the concentration dependence of the EDL capacitance would be directly reflected in the total capacitance. Solvent-separated ion pairing is identified as the key driver of anomalous EDL behaviour in LiCl WiS electrolytes, establishing design considerations for tuning interfacial capacitance and stability in next-generation aqueous energy-storage systems. \end{abstract}

\maketitle
\makeatletter\let\@fnsymbol\orig@fnsymbol\makeatother
\endgroup

\section{Introduction}
At the interface between a charged surface and a liquid electrolyte, ions redistribute and solvent molecules reorient, forming an electrical double layer (EDL) that screens the surface charge. The composition, thickness and structure of the ions in this interfacial layer govern a myriad of processes in environmental, biological, and electrochemical contexts.\cite{shin2022importance,EDLdefinition,simon2020perspectives,rehl2022water, wu2025mechanistic} Early continuum models treated the EDL as a single rigid Helmholtz capacitor. This description was later refined into the Gouy-Chapman-Stern (GCS) model,\cite{stern1924} which assumes that the surface charge is neutralised in two layers: a compact (Stern) layer--locally distinguished into the inner Helmholtz plane (IHP) and outer Helmholtz plane (OHP)--in immediate contact with the surface where counter-ions are strongly adsorbed, and a second, more mobile, diffuse (Gouy-Chapman) layer whose ion density decays towards the bulk solution following the Poisson-Boltzmann (PB) equation. This theoretical framework assumes (i) laterally homogeneous surface charge, (ii) point-like, non-interacting ions, and (iii) a structureless dielectric continuum.

Even at moderate ionic strength, however, the PB assumptions fail: finite ion size, van der Waals, dipolar/solvophobic effects, and strong Coulomb correlations drive specific ion adsorption and complex liquid structuring beyond mean-field theory.\cite{elliott2023specific,howard2010behavior,fedorov2008ionic,laanait2012tuning,wang2001direct, zhan2019specific, hunger2022nature} Extensions that include steric exclusion, dielectric decrement, and ion-ion correlations improve predictions,\cite{borukhov1997steric,fedorov2008ionic,uematsu2018effects, markiewitz2025ionic} but atomistic simulations remain the most direct route to EDL structure.\cite{elliott2022electrochemical,williams_FF, goloviznina2024accounting, jeanmairet2022microscopic} By resolving ion distributions, solvent layering, and interfacial dynamics, molecular dynamics (MD) simulations provide mechanistic access to adsorption free energies, capacitance, and how concentration controls ion accumulation at electrodes.\cite{mceldrew2018theory,Becker2024, li2022unconventional}

MD simulations incorporating explicit surface polarisation have highlighted ion-specific adsorption and nanoscale clustering at graphitic electrodes. Quantum mechanical molecular dynamics (QMMD) simulations by Elliott \textit{et al.} revealed that smaller alkali cations retain their hydration shells and adsorb in the OHP, while larger ions, such as K$^+$, partially dehydrate and penetrate the IHP.\cite{elliott2023specific} Further, they ascribed the asymmetry observed in experimentally measured total specific capacitance to the EDL capacitance ($C_{\mathrm{EDL}}$) contribution, noting that $C_{\mathrm{EDL}}$ is larger at negative potentials because cations have a greater tendency for specific adsorption on graphene. Di Pasquale \textit{et al.}, using constant chemical potential QMMD, demonstrated overscreening and substantial ion clustering in concentrated aqueous alkali halide solutions, a phenomenon traditionally associated with ionic liquids rather than aqueous electrolytes.\cite{di2023constant, vatamanu2017charge, EDLdefinition} More recent QMMD simulations of the graphene/aqueous electrolyte interface show that charging induces an asymmetric restructuring of the interfacial hydrogen‐bond network, with positive potentials markedly increasing water-water hydrogen bonding while negative potentials produce only modest changes.\cite{WeiJACS} Dočkal \textit{et al.}\ showed that charging graphene reorients interfacial water, but individual molecules largely conserve their total number of intermolecular interactions with species by compensatory rearrangements.\cite{dovckal2022molecular} They further demonstrated that larger, weakly hydrated ions adsorb predictably at oppositely charged graphene surfaces, whereas smaller ions can display counterintuitive enrichment (e.g. $\mathrm{Li}^+$ at positively charged interfaces), highlighting the coupled roles of hydration‑shell restructuring and solvent‑mediated correlations in EDL formation. Finney \textit{et al.} found that neutral graphite acquires an effective charge through preferential Na$^+$ adsorption, leading to anion accumulation and, above $\sim$0.6 M, dense cation-anion stacking.\cite{finney2021electrochemistry} Subsequent work demonstrated that negative surface charge amplifies ion clustering and asymmetrically perturbs water orientation.\cite{finney2024properties} These findings highlight how surface polarisation and ion clustering drive rich interfacial structuring not captured by classical continuum models.

Super-concentrated, or “water-in-salt” (WiS), electrolytes push ionic strength to extremes in which solvent molecules become scarce, ion-pairing is enhanced, and conventional solvation shells collapse. These structural features yield wide electrochemical stability windows, suppressed water activity, and unusual transport properties, positioning WiS-based systems as promising electrolytes for high-voltage aqueous batteries and capacitors.\cite{Suo2015Water-in-saltChemistries, borodin2020uncharted, vazquez2025extended} Yet their interfacial structure at carbon electrodes--particularly in alkali-halide WiS electrolytes--and its influence on observables such as capacitance remain largely unexplored.

A central descriptor of interfacial structure is the EDL thickness (or effective screening length, $\lambda_{\mathrm{S}}$).  Classical Debye-Hückel theory predicts that the electrostatic potential decays exponentially over a length $\lambda_{\mathrm{D}}$ -- the Debye length -- and that the effective screening length $\lambda_{\mathrm{S}}\approx\lambda_{\mathrm{D}}$ falls monotonically as salt concentration increases. The GCS model, being grounded in the same PB (and thus Debye-Hückel) mean‑field framework, likewise predicts an exponentially decaying potential and a screening length that falls with increasing ionic strength.  Modern liquid‑state theories that embed finite‑ion size and correlations instead predict a minimum  $\lambda_{\mathrm{S}}$ at the so-called Kirkwood line (the bulk concentration at which charge-charge correlations transition from monotonic to damped-oscillatory decay) followed by a modest increase at higher concentrations.\cite{attard1993asymptotic, adar2019screening,zeman2020bulk, zeman2021ionic} Force‑measurement studies on concentrated electrolytes, however, report much larger values that grow approximately as $\lambda_{\mathrm{D}}^{3}$ and extend tens of nanometres,\cite{lee2017scaling} a phenomenon termed “underscreening”. Consequently, the value of the screening length in alkali‑halide electrolytes with concentration remains an open and actively debated question.

In the present work, we use classical MD and QMMD to simulate aqueous LiCl from moderate to WiS concentrations at neutral and charged graphitic electrodes. Novel scaled-charge force field models have facilitated the investigation of much higher concentration regimes than were previously accessible. From these simulations, we study the concentration-dependent EDL structure and determine both the EDL and quantum capacitances. Our simulations show a significant transition in the EDL structure and thickness on graphite, occurring above $4~\mathrm{mol~kg^{-1}}$ for neutral surfaces and at even lower concentrations when charged. This transition is driven by the preferential adsorption of $\mathrm{Li}^+$ and substantial co-adsorption of $\mathrm{Cl}^-$ into the OHP via solvent-separated ion pairs. This co-adsorption in the OHP does not fully neutralise the electrode charge, leaving a residual field that drives alternating ion layers and expands the effective EDL thickness. Consequently, the potential of zero charge (PZC) decreases, the interfacial potential drop redistributes, and the $C_{\mathrm{EDL}}$ changes substantially, with direct consequences for electrochemical performance and electrolyte design. We validate the simulated concentration dependence of the PZC against values extracted from differential-capacitance measurements obtained via electrochemical impedance spectroscopy.

\section{Methodology}
\subsection{Computational Model} \label{model}
\begin{figure}[htbp]
  \centering
  \includegraphics[
    width=\textwidth
  ]{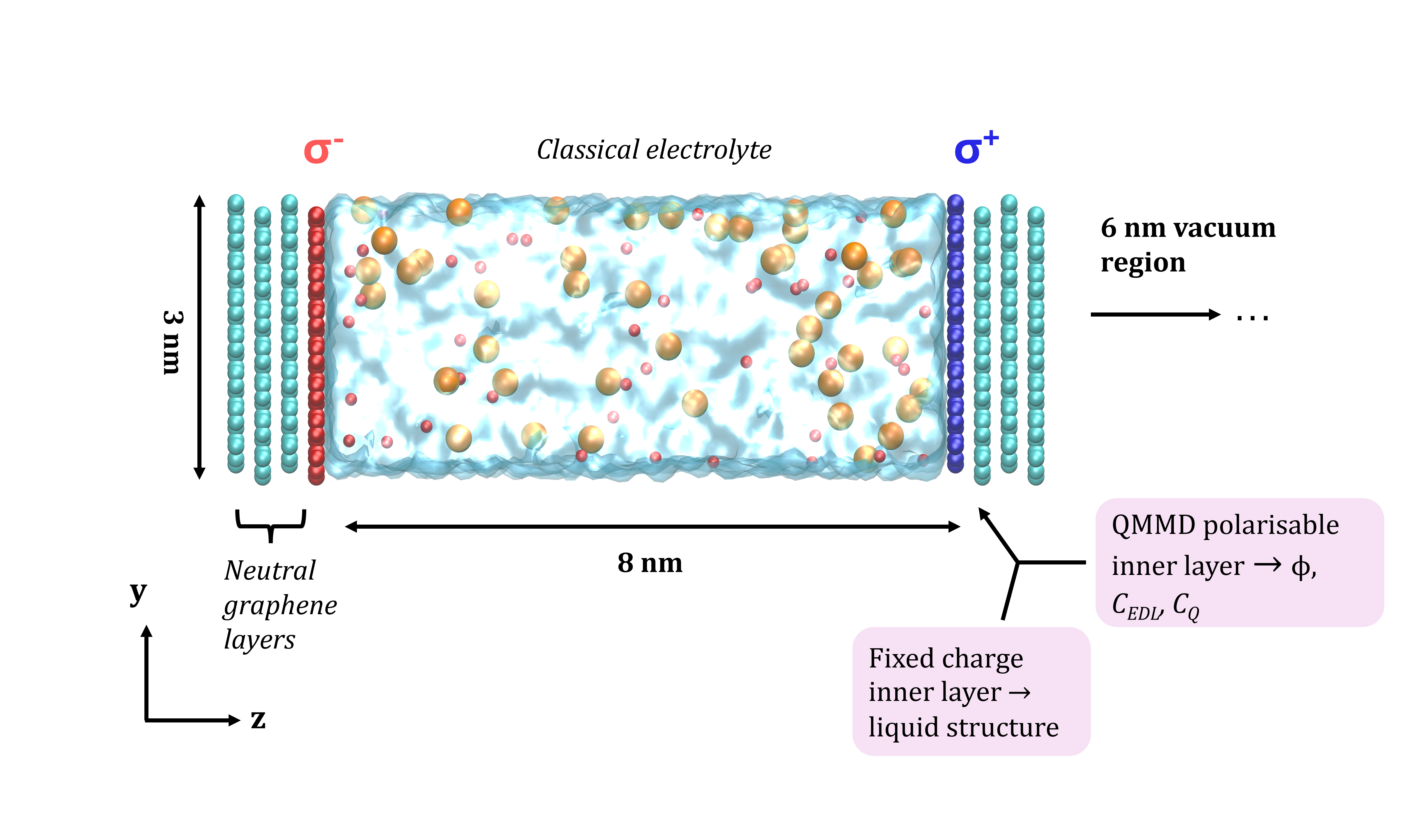}
  \caption{Schematic of the simulation cell showing a liquid electrolyte confined between two graphite electrodes in the $x$-$y$ plane. Each electrode consists of four stacked graphene sheets; the innermost sheet carries a surface charge density of $\sigma^0$= 0 $\mathrm{C~m}^{-2}$ for neutral systems, or $\sigma^+$ = +0.0571 $\mathrm{C~m}^{-2}$ (blue) and $\sigma^-$ = -0.0571 $\mathrm{C~m}^{-2}$ (red) in charged systems, while the three subsurface sheets remain neutral (cyan).}
  \label{fig:simulation_cell}
\end{figure}
We model a parallel-plate capacitor configuration, comprising a $\mathrm{LiCl}$ slab ($3\times3\times8~\mathrm{nm}^3$) confined between two graphite electrodes, each built from four stacked graphene sheets ($3\times3~\mathrm{nm}^2$); see Figure~\ref{fig:simulation_cell}. A $6~\mathrm{nm}$ vacuum gap along the surface normal was imposed to prevent interactions with periodic images.\cite{longrangeffect} We consider three surface-charge states, $\sigma^{0}=0$, $\sigma^{-}=-0.0571$, and $\sigma^{+}=+0.0571~\mathrm{C~m^{-2}}$, applying charge only to the inner graphene layers in contact with the electrolyte. $\mathrm{LiCl}$ concentrations span $1$--$20~\mathrm{mol~kg^{-1}}$ (1, 2, 4, 6, 10, 16, 20). Following Ref.~\citenum{WeiJACS}, we use the Madrid-2019 ion force field,\cite{Mardid2019} TIP4P/2005 water constrained with SETTLE,\cite{Tip4p2005} and take LJ parameters from Cornell \textit{et~al.}\cite{AmberCarbon} for the graphitic carbons. The C-O Lennard-Jones $\varepsilon$ parameter is taken from Werder \textit{et~al.} to reproduce the experimental water/graphite contact angle,\cite{Werder,Loischannelpaper} and ion-carbon non-bonded terms use Lorentz-Berthelot combining rules.

\subsection{Classical MD}\label{classical}
Classical MD simulations were performed with the GROMACS 2018.4 software package.\cite{GROMACS1,GROMACS2} Numbers of water molecules and ions were selected to reproduce the desired concentration and bulk density, as computed in our previous work,\cite{HannahPaper} with further details in the supporting information (SI). All simulations were carried out in the NVT ensemble with a Nosé-Hoover thermostat at 294.5~K (relaxation time 0.5~ps). The real-space cutoff for Lennard-Jones and Coulomb interactions was 1.2~nm. Long-range electrostatics were treated with particle-mesh Ewald (standard 3D geometry). The Lennard-Jones 12--6 potential used a switching function from 1.0 to 1.2~nm to smooth the truncation. Graphite electrode atoms are frozen in space during the simulation. Trajectories for $1$--$6~\mathrm{mol~kg^{-1}}$ ran $450~\mathrm{ns}$, whereas those above $6~\mathrm{mol~kg^{-1}}$ were extended to $750~\mathrm{ns}$ to obtain more statistically robust data. This extension is justified by our previous work, which showed that $\mathrm{LiCl}$ solutions with high concentrations require longer simulation times to reach thermodynamic equilibrium.\cite{HannahPaper}  

\subsection{QMMD}\label{QMMD}
For the calculation of total capacitance, we simulated the capacitor introduced in previous section with the QMMD scheme of Elliott \textit{et al.}\cite{JoshPaper} In this iterative method, the electronic structure of the innermost graphite layer is updated by self-consistent-charge density-functional tight binding (SCC-DFTB) calculations, while the positions of the electrolyte atoms are propagated by classical MD (with frozen electrode carbon atoms). At each quantum step, the SCC-DFTB Hamiltonian incorporates the instantaneous electrostatic field produced by the MD point charges; a Mulliken population analysis maps the resulting charge distribution onto partial charges of the carbon atoms, enabling the electrode to polarise self-consistently in response to its ionic environment. Quantum and classical subsystems are coupled every 5 ps: after each 5 ps segment of MD, a new SCC-DFTB step is performed to refresh the carbon charges, a coupling interval validated in previous works.\cite{JoshPaper} 

The MD parameters employed in QMMD calculations match those of the fully classical simulations described in classical MD section. Electronic-structure updates were performed with the \textsc{DFTB+} software package,\cite{hourahine2020dftb+} with the \texttt{mio-1-1} parameter set for C-C interactions.\cite{elstner1998self} Calculations were restricted to the $\Gamma$-point with a Fermi smearing of $1\times10^{-6}~\mathrm{K}$. The self-consistent-charge cycle was considered converged when the total energy changed by less than $1\times10^{-2}~\mathrm{Ha}$. These settings have been shown to reproduce the surface charge distribution obtained from higher-level quantum-mechanical calculations within statistical uncertainty.\cite{JoshPaper, elliott2023specific} Each system was first equilibrated for 60 ns under NVT conditions, followed by a 30 ns production run with the QMMD scheme.

\subsection{Experimental methods}
Capacitance measurements were conducted on the basal plane of highly oriented pyrolytic graphite (HOPG) using electrochemical impedance spectroscopy (EIS) over a frequency range of $20~\mathrm{kHz}$--$1~\mathrm{Hz}$ with a $7~\mathrm{mV}$ rms amplitude. The electrochemical cell configuration used a polytetrafluoroethylene cylinder with a disk-shaped opening of $3~\mathrm{mm}$. A poly(dimethylsiloxane) gel layer (Sylgard\textsuperscript{TM} 527, Dow Corning) was used to seal between the bottom of the PTFE cylinder and the HOPG to prevent solution leakage. The application of potentials for EIS starts from $0~\mathrm{V}$ (vs $\mathrm{Ag/AgCl}$ in $3.5~\mathrm{M}$ KCl), with $50~\mathrm{mV}$ potential steps, alternating between positive and negative potentials.

The effective capacitance ($C_{\mathrm{eff}}$) at each frequency was then calculated from the EIS data by adopting a method developed by Orazem and colleagues,\cite{orazem2006enhanced}
\begin{equation}
  C_{\mathrm{eff}}
  = -\sin\left(\frac{\alpha\pi}{2}\right)
    \frac{1}{Z_{\mathrm{im}}\bigl(2\pi f\bigr)^{\alpha}},
\end{equation}
where $Z_{\mathrm{im}}$ is the imaginary part of the impedance and $f$ is the frequency. The parameter $\alpha$ is the constant-phase-element exponent, which can be estimated from the slope between $\log f$ and $\log Z_{\mathrm{im}}$. The final capacitance values were determined by averaging $C_{\mathrm{eff}}$ from the frequency range of 10 Hz to 100 Hz. 
\section{Results and Discussion}
\subsection{Electrode/Electrolyte Interface Structure }\label{sec:structure}
To investigate the interfacial liquid structure, we first calculated the bulk-normalised number density $\tilde{n}(z)$ of ionic and solvent species along the surface-normal ($z$) direction, 
\begin{equation}
  \tilde{n}(z)= \frac{\rho(z)}{\rho_{bulk}},
\end{equation}
where $\rho_{bulk}$ represents the number density of each species in the bulk region.

The near-surface region contains two characteristic planes--the IHP and the OHP--beyond which the density profiles decay in the diffuse layer (DL) towards bulk-like behaviour. The IHP is defined as the plane passing through the centre of the first adsorbed layer of species (predominantly water) and is therefore located at the maximum of the first peak in the water-oxygen density profile. Analogously, the OHP is defined as the plane through the centres of the second layer of species (first layer of solvated ions) and is therefore taken as the position of the first $\mathrm{Li}^+$ peak, since $\mathrm{Li}^+$ adsorption is observed at all surface charges and concentrations and occurs predominantly in the OHP. It is well established that cations in alkali chloride electrolytes reside closer to neutral graphitic surfaces than chloride anions.\cite{elliott2023specific, williams_FF, finney2021electrochemistry, dockal2019molecular} This reflects a free-energy balance involving favourable cation-surface interactions (captured explicitly as cation-$\pi$ polarisation in models including surface polarisability, or effectively through the Lennard-Jones cross terms in simpler models), hydration free energies, and entropic packing constraints. Previous studies have reported that neglecting cation-$\pi$ interactions can underestimate cation adsorption;\cite{elliott2023specific} however, in our systems, comparison of QMMD (including surface polarisation) with classical MD showed no such underestimation. This may be due to our use of the scaled-charge Madrid-2019 ion model, which has been shown to reproduce interfacial properties of alkali cations. Despite the absence of imposed surface charge density, this specific cation adsorption produces a nonzero PZC, which can be regarded as an effective surface charge that renders a structured EDL in much the same way as an explicitly charged electrode. 

\begin{figure}[htbp]
  \hspace*{-0.07\textwidth} 
  \begin{overpic}[width=1.1\textwidth]{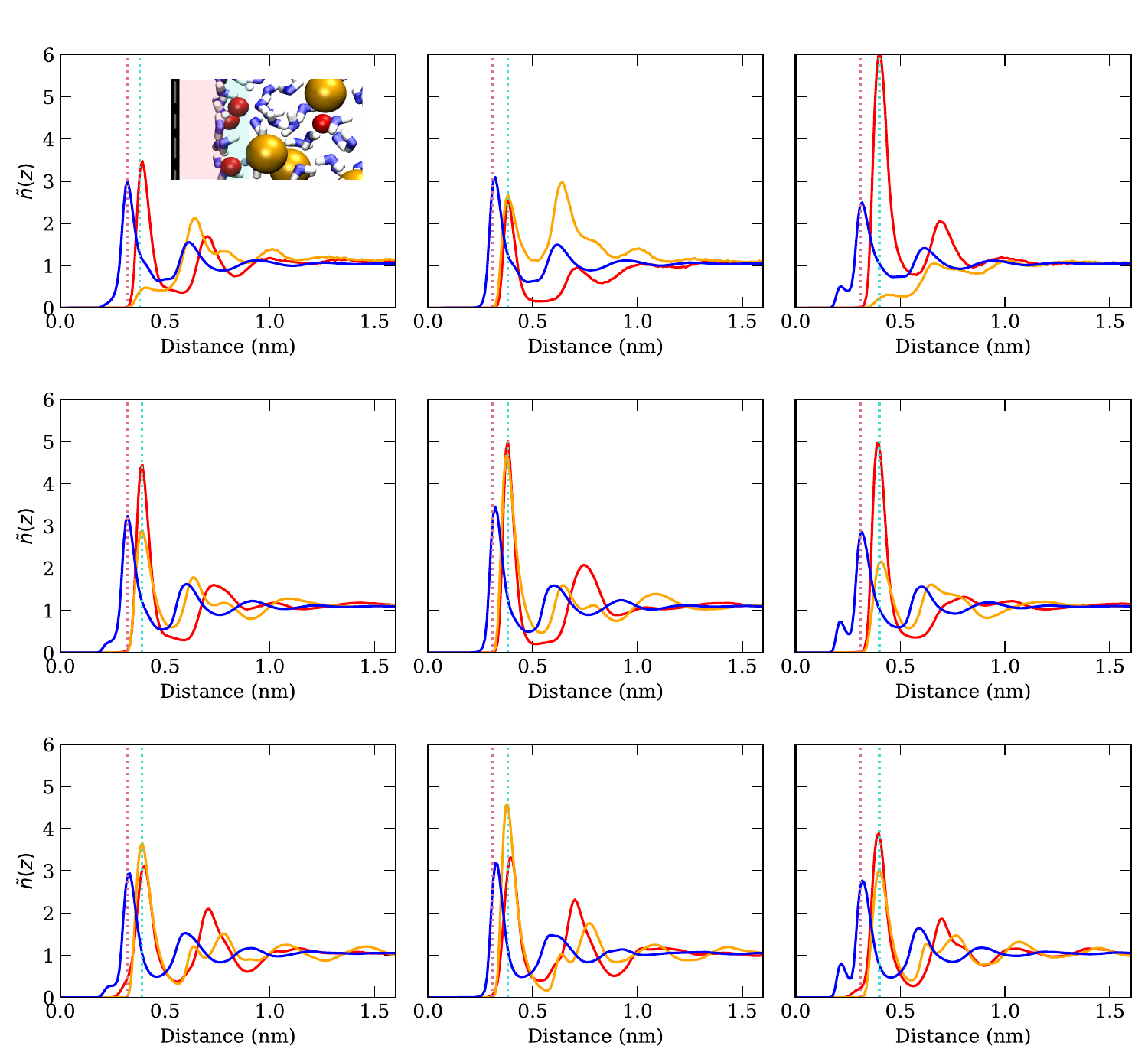}
    \put(5,89){\textbf{(a)}}
    \put(37,89){\textbf{(b)}}
    \put(69,89){\textbf{(c)}}
    \put(5,59){\textbf{(d)}}
    \put(37,59){\textbf{(e)}}
    \put(69,59){\textbf{(f)}}
    \put(5,29){\textbf{(g)}}
    \put(37,29){\textbf{(h)}}
    \put(69,29){\textbf{(i)}}
    \put(20,89){\large\textbf{$\sigma^0$}}
    \put(52,89){\large\textbf{$\sigma^+$}}
    \put(84,89){\large\textbf{$\sigma^-$}}   
  \end{overpic}
\caption{Bulk-normalised number-density profiles, $\tilde n(z)$, for $\mathrm{Li}^+$ (red), $\mathrm{Cl}^-$ (orange), and water atoms (blue) as a function of distance $z$ from the electrode surface. Columns correspond to electrode surface charge densities $\sigma^0=0~\mathrm{C~m^{-2}}$ (left), $\sigma^+=+0.0571~\mathrm{C~m^{-2}}$ (middle), and $\sigma^-=-0.0571~\mathrm{C~m^{-2}}$ (right). Rows show concentrations: (a--c) $1~\mathrm{mol~kg^{-1}}$ LiCl; (d--f) $10~\mathrm{mol~kg^{-1}}$; (g--i) $20~\mathrm{mol~kg^{-1}}$. Vertical dotted lines indicate the inner Helmholtz plane (IHP, pink) and outer Helmholtz plane (OHP, cyan).}

  \label{fig:dens_profile}
\end{figure}
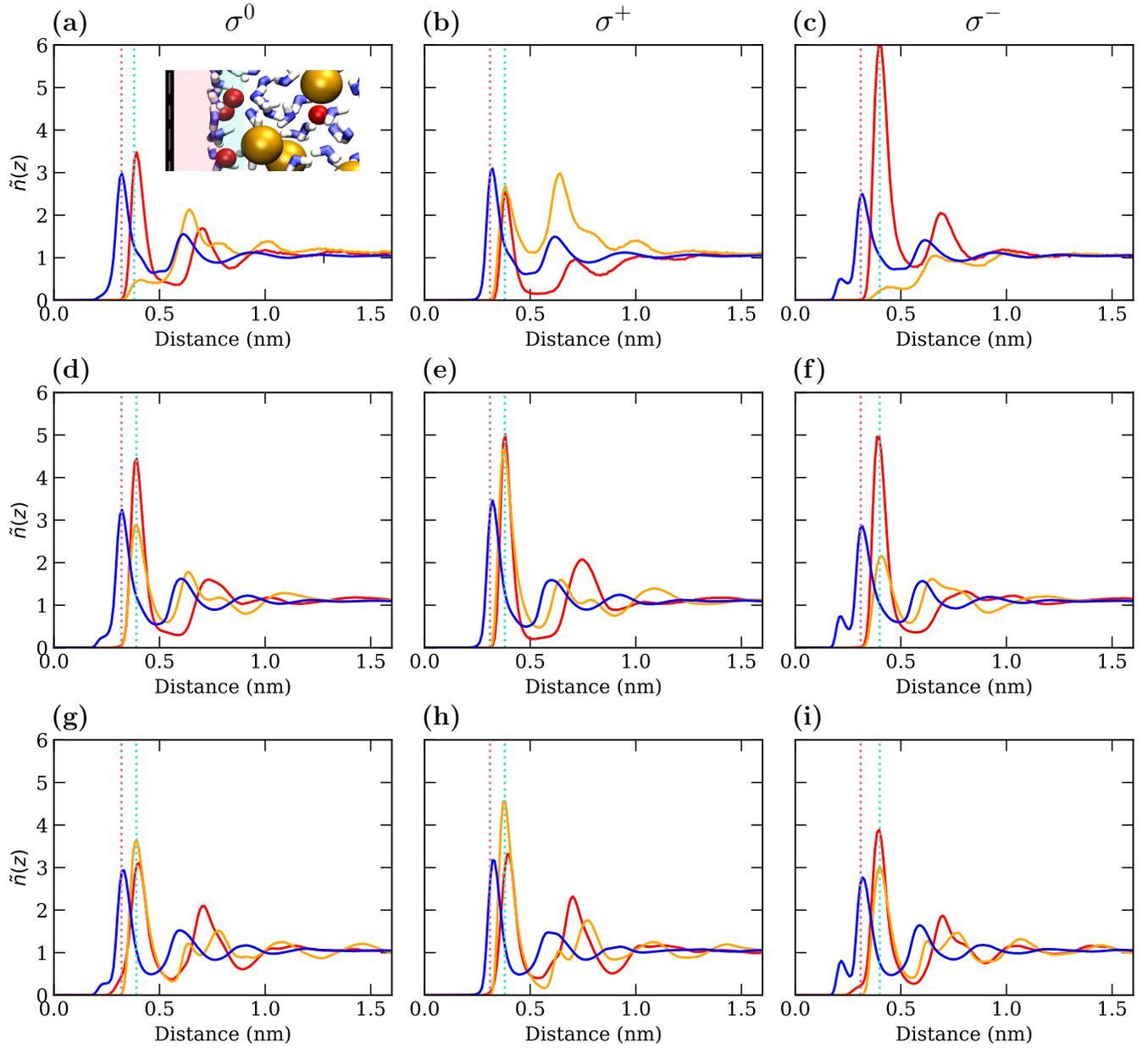

The positions of the IHP and OHP were determined for each system. Although the absolute positions of the IHP and OHP shift slightly with surface charge, they are essentially insensitive to concentration for each charge state. In all distance-resolved plots, we mark the IHP and OHP with pink and cyan dotted lines, respectively, and a representative snapshot highlighting these planes is included in Figure~\ref{fig:dens_profile} (a). Figure~\ref{fig:dens_profile} shows $\tilde{n}(z)$ profiles of $\mathrm{LiCl}$ electrolytes at $1~\mathrm{mol~kg^{-1}}$, $10~\mathrm{mol~kg^{-1}}$, and $20~\mathrm{mol~kg^{-1}}$ under different surface charges. Density profiles for other concentrations are provided in the SI.

For all non-charged systems, we observe that water molecules form layers near the graphite surface. The first water layer is located at approximately 0.32~nm from the graphite at the IHP, while the second layer appears around 0.60~nm. This interfacial structuring is consistent with previous studies\cite{graphite-water} on the graphite-water interface using the TIP4P/2005 water model. In all neutral systems, the first peak of $\mathrm{Li}^+$ appears approximately 0.39~nm from the surface, followed by a second peak around 0.70~nm. For $\mathrm{Cl}^-$, the first and second peaks are located at approximately 0.39~nm and 0.64~nm, respectively. 

At concentrations below $6~\mathrm{mol~kg^{-1}}$, we observe the clear adsorption of $\mathrm{Li}^+$ in the OHP at neutral graphite, with $\mathrm{Cl}^-$ predominantly residing farther from the surface.\cite{WeiJACS,elliott2023specific} As the concentration of $\mathrm{LiCl}$ reaches $10~\mathrm{mol~kg^{-1}}$, significant changes are observed in the distribution of $\mathrm{Cl}^-$ ions: anions enter the OHP, accumulating at the same location as the $\mathrm{Li}^+$ (shown in Figure~\ref{fig:dens_profile} (d), where the $\mathrm{Cl}^-$ density profile first peak increases in height, surpassing that of its second peak). At 1$6~\mathrm{mol~kg^{-1}}$, this accumulation of $\mathrm{Cl}^-$ becomes even more evident, and the peak intensity matches that of $\mathrm{Li}^+$ (Figure S6 (a)). When the concentration reaches $20~\mathrm{mol~kg^{-1}}$ (Figure~\ref{fig:dens_profile} (g)), the OHP $\mathrm{Cl}^-$ peak height exceeds that of $\mathrm{Li}^+$. At the same time, a small fraction of $\mathrm{Li}^+$ extends into the IHP, as shown by the weak density tail beyond the pink dotted line, which lowers the proportion remaining in the OHP and decreases the $\mathrm{Li}^+$ peak intensity.

The concentration-dependent redistribution of $\mathrm{Cl}^-$ ions becomes more pronounced once the electrode is charged. Figure~S8 presents bulk-normalised density profiles of $\mathrm{Li}^+$ and $\mathrm{Cl}^-$ across the full concentration range. At the positively charged electrode, $\mathrm{Cl}^-$ adsorption strengthens systematically: the accumulation shift evident in the neutral case already appears at $2~\mathrm{mol~kg^{-1}}$, the first peak height becomes comparable to that of $\mathrm{Li}^+$ by $10~\mathrm{mol~kg^{-1}}$, and surpasses it at $16~\mathrm{mol~kg^{-1}}$. On the negative electrode, $\mathrm{Cl}^-$ adsorption is weakened and shifted slightly outward relative to the neutral case, though appreciable density persists in the OHP at concentrations above $10~\mathrm{mol~kg^{-1}}$, never exceeding that of $\mathrm{Li}^+$. In both polarities the water oxygen peaks retain their positions, although reorientation of the molecules occurs in response to the electrode charge.  

Between 4 and $10~\mathrm{mol~kg^{-1}}$, positive charging yields an enhanced $\mathrm{Li}^+$ population in the OHP despite the expected electrostatic repulsion, with adsorption levels comparable to those at the negatively charged electrode and higher than at the neutral interface. At $20~\mathrm{mol~kg^{-1}}$, positive charging suppresses the small $\mathrm{Li}^+$ population in the IHP while leaving the OHP density essentially unchanged, whereas negative charging slightly increases the IHP density. Overall, $\mathrm{Cl}^-$ responds to electrode polarity in the expected manner-enhanced at positive, reduced at negative potentials--while $\mathrm{Li}^+$ shows a weaker and in some cases counterintuitive response, particularly the OHP enhancement at the positively charged electrode in the $4-10~\mathrm{mol~kg^{-1}}$ range (\hyperref[sec:driving_force]{discussed further in the analysis of the local charge density}).

\subsection{Electric Double Layer Thickness}
The thickness of the EDL at the electrode/electrolyte interface plays a critical role in determining electrochemical behaviour, and has been studied extensively in the literature.\cite{EDLmeasure1,EDLmeasure2,EDLmeasure3,EDLmeasure4, guerrero2018quantifying} Traditionally, it is estimated using the Debye length, which assumes point-like ions and a dilute electrolyte under the PB framework and can be calculated from molecular simulations as the distance at which the cumulative ionic charge density $ Q(z) $ reaches a plateau\cite{EDLmeasure2} or where the ionic profiles return to bulk-like values.\cite{EDLmeasure3}

\begin{figure}[H]
  \centering
  \begin{overpic}[width=0.9\textwidth]{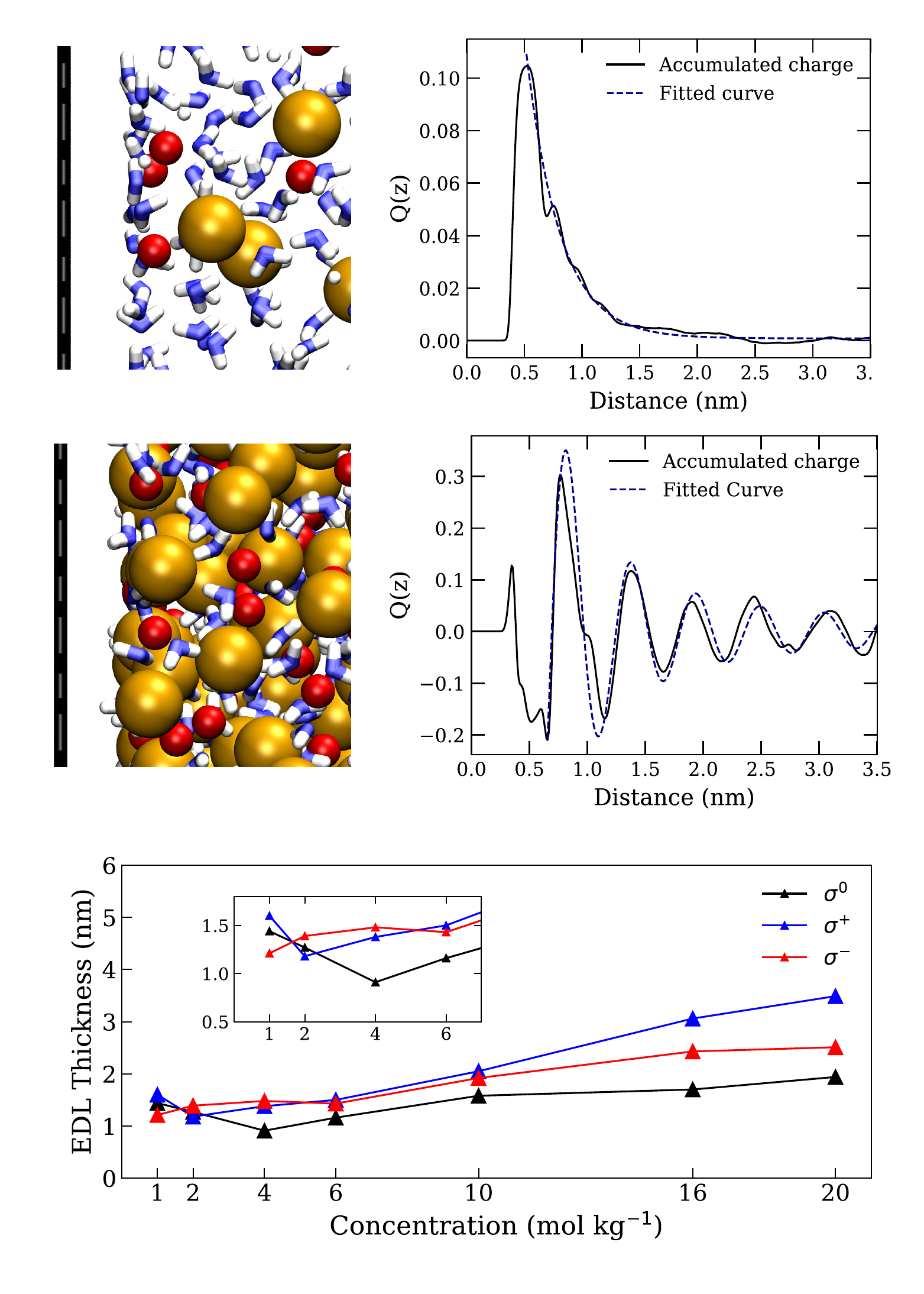}
    \put(0,95){\textbf{(a)}}
    \put(0,64){\textbf{(b)}}
    \put(29,95){\textbf{(c)}}
    \put(29,64){\textbf{(d)}}
    \put(0,35){\textbf{(e)}}
  
  \end{overpic}
  \caption{
     Electric double layer thickness estimation. (a, b) Snapshots of the interfacial region for $1~\mathrm{mol~kg^{-1}}$ and $20~\mathrm{mol~kg^{-1}}$ LiCl in contact with neutral graphite. (c, d) Accumulated charge density profiles $Q(z)$ (black solid line) near the neutral surface, together with fitted damped oscillatory curves (blue dashed line), corresponding to $1~\mathrm{mol~kg^{-1}}$ and $20~\mathrm{mol~kg^{-1}}$ LiCl. (e) EDL thickness as a function of electrolyte concentration under three surface charge conditions: $\sigma^{0}$ (black), $\sigma^{+}$ (blue), and $\sigma^{-}$ (red).} 
  \label{fig:EDL} 
\end{figure}
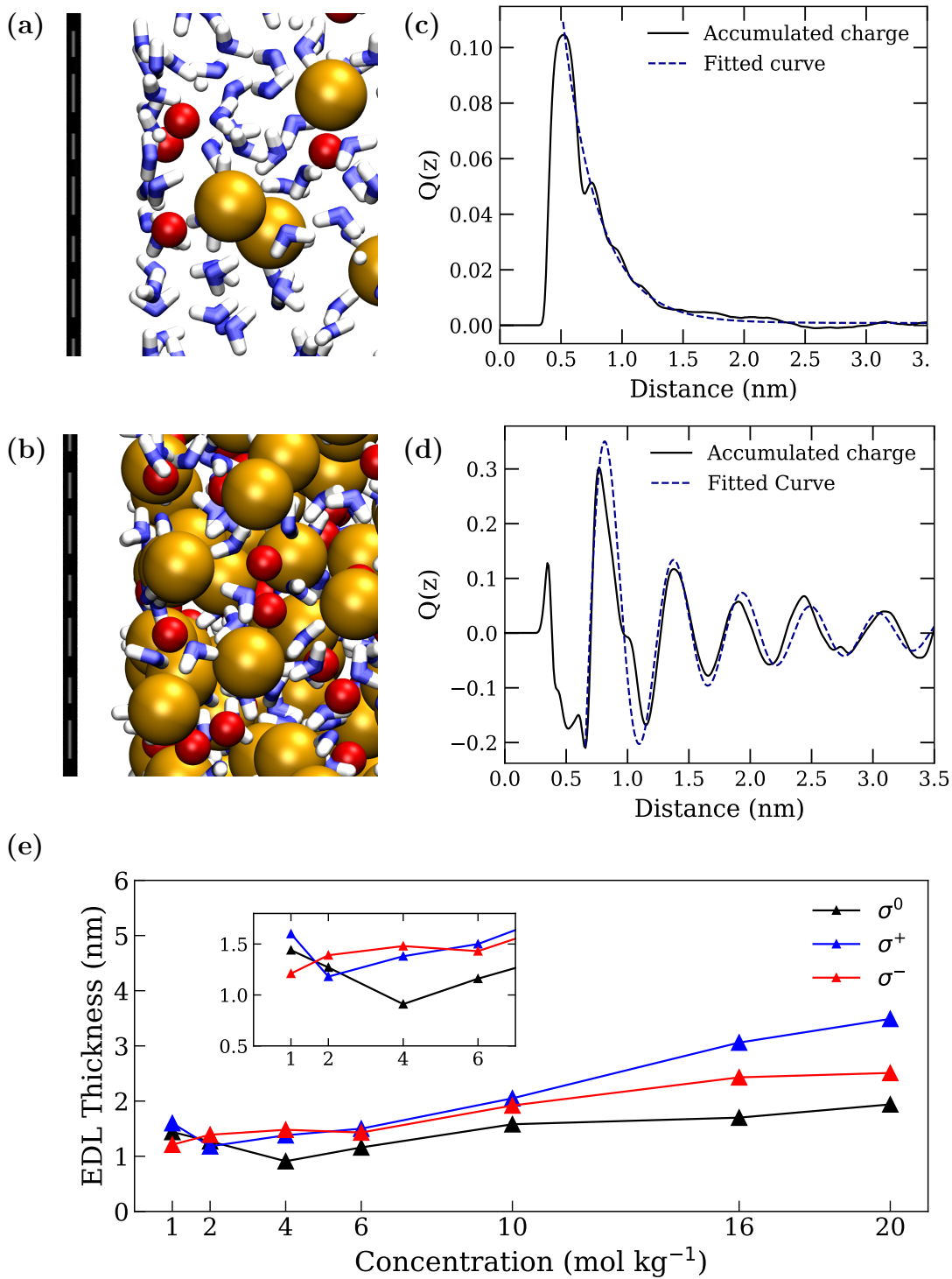

In this study, the thickness of the EDL is determined using a two-step approach: First, computing the accumulated charge density $ Q(z) $ in the neutral system and the screening factor $S(z)$ in charged systems,
\begin{equation}
Q(z) = \int_0^z \rho_{\mathrm q}^{\mathrm{ion}}(z') ~ \mathrm{d}z'
\end{equation}
\begin{equation}
S(z) = \frac{1}{\sigma} \int_0^z \rho_{\mathrm q}^{\mathrm{ion}}(z') ~ \mathrm{d}z'
\end{equation}
Here, $\rho_{\mathrm q}^{\mathrm{ion}}(z')$ denotes the net ionic charge density as a function of the distance from the electrode, $z$, and $ \sigma $ is the surface charge of the electrode. As both $Q(z)$  and $S(z)$ show oscillatory behaviour, the resulting curves are fitted by an under-damped oscillation function,
\begin{equation}
f(z) = A(z) \cdot \cos(\omega z + \phi) + C
\end{equation}
where $ A(z) $ is an envelope function for the amplitude, $ \omega $ is the angular frequency, $ \phi $ is the phase change and $ C $ is the baseline offset. This approach has been applied previously to describe the decaying, oscillatory screening at electrified interfaces with concentrated electrolytes and ionic liquids.\cite{jager2023screening, EDLmeasure2}  For concentrations ranging from 1 to $10~\mathrm{mol~kg^{-1}}$, the accumulated charge profiles were fitted using a standard damped cosine model with a constant damping coefficient:
\begin{equation}
f(z) = A \cdot e^{-k z} \cdot \cos(\omega z + \phi) + C
\end{equation}
where $A e^{-k z}$ represents the exponential decay envelope. However, at higher concentrations ($16-20~\mathrm{mol~kg^{-1}}$), the oscillatory behaviour of $Q(z)$ (or $S(z)$) becomes significantly more pronounced, and the simple exponential damping fails to capture the persistence of oscillations. To address this, a variable damping model was employed, in which the damping coefficient decreases with distance,
\begin{equation}
f(z) = A \cdot e^{\left(-\frac{k_0}{1+z} \cdot z\right) }\cdot \cos(\omega z + \phi) + C .
\end{equation}
Here, the damping rate $k(z) = \frac{k_0}{1 + z}$ decreases with increasing $z$, allowing the fit to better account for the extended range and persistence of oscillations observed in high-concentration regimes. The EDL thickness is defined as the position $ z = z^* $ where the deviation of the envelope function $ \Delta A(z) $ falls below 5\% of the maximum of the accumulated charge or screening factor, i.e.,
\begin{equation}
\Delta A(z^*) \leq 0.05 \cdot \max \left\{ Q(z),~ S(z) \right\}.
\end{equation}
Figure~\ref{fig:EDL} (a) and (b) show representative snapshots of the electrolyte structure at the neutral graphite interface under concentrations of $1~\mathrm{mol~kg^{-1}}$ and $20~\mathrm{mol~kg^{-1}}$, respectively. The corresponding accumulated charge density profiles $Q(z)$ are presented in Figure~\ref{fig:EDL} (c) and (d), together with the fitted damped oscillatory curves. The results of the estimated thickness of EDL are summarized in Figure~\ref{fig:EDL} (e) as a function of electrolyte concentration under three surface charge conditions. Further details on the thicknesses of individual layers are provided in the SI.

In Debye-Hückel theory, EDL thickness decreases monotonically with concentration as additional ions are available to screen the surface charge; our data show a more complex trend. For the neutral case, the EDL thickness first decreases and then increases as the concentration of $\mathrm{LiCl}$ rises. We attribute this trend reversal to the enhanced oscillatory behaviour of ions at high concentrations. Although the thickness of the IHP and OHP remains relatively unchanged, the diffuse layer is significantly extended due to intensified ionic layering, thereby increasing the overall EDL thickness.

Under positively charged conditions, a similar non-monotonic trend is observed, with EDL thickness decreasing from $1-2~\mathrm{mol~kg^{-1}}$, but increasing beyond $2 ~\mathrm{mol~kg^{-1}}$: the onset of increasing thickness occurs at a lower ionic concentration compared with the neutral case. This is consistent with the earlier accumulation of $\mathrm{Cl}^-$ in the OHP, which induces overscreening and extends the oscillatory tail of $S(z)$, increasing the fitted EDL thickness.

In contrast, for negatively charged surfaces, the EDL thickness increases monotonically with concentration. Below $6~\mathrm{mol~kg^{-1}}$ the EDL thickness changes only marginally. Although more $\mathrm{Li}^+$ ions adsorb at the electrode at higher concentrations, the negative surface potential repels $\mathrm{Cl}^-$ out of the OHP. This reduces their capacity to counterbalance the excess $\mathrm{Li}^+$ accumulation. Once the concentration exceeds $10~\mathrm{mol~kg^{-1}}$, a clear accumulation shift of $\mathrm{Cl}^-$ ions is observed, resulting in an expansion of the diffuse layer.

\subsection{Driving forces leading to the interfacial structural change}\label{sec:driving_force}
\begin{figure}[htbp]
  \centering
  \includegraphics[
    width=\textwidth,
    height=0.72\textheight,
    keepaspectratio
  ]{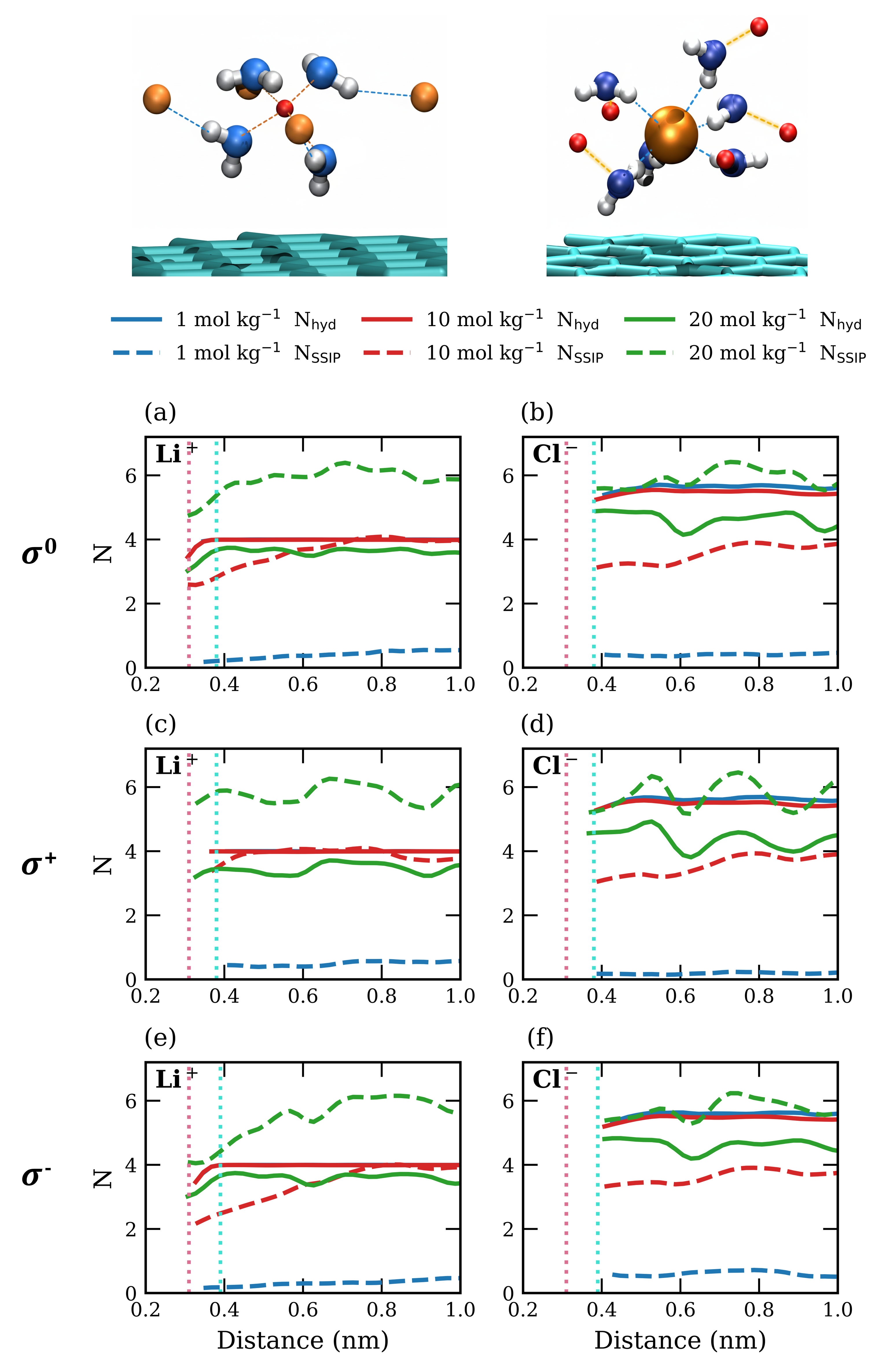}
  \caption{Hydration number, $N_{\mathrm{hyd}}$, calculated as the mean number of H$_2$O-ion intermolecular bonds per ion (solid lines), and solvent-separated ion pairs per ion, $N_{\mathrm{SSIP}}$ (dashed lines), as a function of distance from the electrode for $\mathrm{Li^{+}}$ (left column) and $\mathrm{Cl^{-}}$ (right column). Concentrations of 1, 10, and $20~\mathrm{mol~kg^{-1}}$ are shown in blue, red, and green, respectively. Vertical dotted lines mark the inner Helmholtz plane (IHP, pink) and the outer Helmholtz plane (OHP, cyan). Row labels indicate the electrode surface charge: $\sigma^{0}$ (top row, neutral), $\sigma^+=+0.0571~\mathrm{C~m^{-2}}$ (middle row), and $\sigma^-=-0.0571~\mathrm{C~m^{-2}}$ (bottom row). The images above depict representative hydration motifs for $\mathrm{Li^{+}}$ (left) and $\mathrm{Cl^{-}}$ (right).
}
  \label{fig:CN}
\end{figure}

To understand the driving forces governing the concentration-induced restructuring of the EDL, we analyse the ions' local chemical environment using the procedure introduced by Do\v{c}kal and coworkers,\cite{dovckal2019general} which uses topological features of the spatial distribution function to quantify properties such as hydration numbers and ion pairing. As the method does not rely on pre-defined distance-angle cut-offs, it enables transferable, non-redundant quantification of the local molecular arrangement, which is particularly important in the context of WiS electrolytes, where liquid structure deviates substantially from the dilute regimes for which conventional cut-off-based criteria were developed. The method works by defining "intermolecular bonds" using the spatial distribution function (SDF), which gives the relative probability of finding a monitored particle (in our case O, H, $\mathrm{Li}^+$, or $\mathrm{Cl}^-$) at a position $(x, y, z)$ relative to a reference (water) molecule. Regions of high probability in the SDF (local maxima) indicate favoured positions around the monitored species. An intermolecular bond is considered to exist when the particle is located within a continuous volumetric region surrounding such a maximum; this volume is bounded by the isosurface that intersects the nearest "significant" saddle point (NSSP).\cite{dovckal2019general} This approach ensures that only particles in well-defined, high-probability regions, corresponding to physically meaningful interactions, are counted as bonded. Using these bonding criteria, we determine both the ion-water hydration number, $N_{\mathrm{hyd}}$, and the number of solvent-separated ion pairs (SSIPs), $N_{\mathrm{SSIP}}$, which we identify as pairs of oppositely charged ions that are bonded to a common water molecule. The $z$-resolved results for neutral graphite systems are presented in Figure~\ref{fig:CN} (a, b). Across all $z$ and for both $\sigma^{+}$ and $\sigma^{-}$ electrodes, no contact ion pairs were detected for LiCl concentrations $\le 10 ~\mathrm{mol~kg^{-1}}$;\cite{dovckal2024structure}  at $16$ and $20 ~\mathrm{mol~kg^{-1}}$ only a minor contact ion pair population appears, with mean Li coordination numbers (over all $z$) of $N_{\mathrm{Li-H_2O}}=3.9$ and $N_{\mathrm{Li-Cl}}=0.1$ at $16~\mathrm{mol~kg^{-1}}$, and $3.5$ and $0.5$ at $20~\mathrm{mol~kg^{-1}}$, respectively. The hydration number of $\mathrm{Li}^+$ is essentially concentration-independent: it remains $N_{\mathrm{hyd}}\simeq4$ at 1 and $10~\mathrm{mol~kg^{-1}}$ and decreases only slightly to $\sim3.5$ at $20~\mathrm{mol~kg^{-1}}$. This small drop reflects the scarcity of free water in the WiS regime.

Throughout the bulk liquid, diffuse layer, and OHP, the $\mathrm{Li}^+$ hydration number $N_{\mathrm{hyd}}$ is essentially independent of $z$. Figure~\ref{fig:dens_profile} reveals that a small fraction of $\mathrm{Li}^+$ ions penetrate into the IHP--most conspicuously at $20~\mathrm{mol~kg^{-1}}$--as evidenced by the tail of the first cation density peak into this region. Once inside the IHP, $N_{\mathrm{hyd}}$ decreases to $\sim 3$, while the $\mathrm{Li}^+$ in the OHP retain their full primary hydration shell.\cite{elliott2023specific} Chloride behaves differently: the anions never enter the IHP, and their hydration number drops from $\sim5.8$ at 1 and $10~\mathrm{mol~kg^{-1}}$, reflecting a mixture of typical penta-/hexa-coordinated configurations, to $\sim4.8$ at $20~\mathrm{mol~kg^{-1}}$. Because $\mathrm{Cl}^-$ is larger and more weakly hydrated, it loses water molecules more readily in the WiS environment.  

The number of SSIPs rises sharply with concentration, from $\approx0.5$ per $\mathrm{Li}^+$ at $1~\mathrm{mol~kg^{-1}}$ to $\sim3.8$ at $10~\mathrm{mol~kg^{-1}}$ and $\sim5.8$ at $20~\mathrm{mol~kg^{-1}}$. SSIPs are slightly less abundant in the OHP and IHP than in the diffuse and bulk regions, decreasing to approximately 3.0 per Li at $10~\mathrm{mol~kg^{-1}}$ and 5.4 per Li at $20~\mathrm{mol~kg^{-1}}$. A modest suppression of SSIPs in the first ion adsorption layer is unsurprising, where a rigid surface creates geometric restriction. 

These results indicate that the equal densities of $\mathrm{Li}^+$ and $\mathrm{Cl}^-$ in the OHP of the WiS solutions arise because $\mathrm{Li}^+$ retains a tightly bound first hydration shell, even at high concentrations. Each $\mathrm{Li}^+$ remains coordinated to $\sim4$ water molecules across all salt concentrations, thanks to its small radius and strongly negative hydration free energy.\cite{ovalle2022correlating} Therefore, $\mathrm{Li}^+$-bound water molecules are left with two H-donor sites that must be satisfied; in dilute solutions, they hydrogen-bond to other waters, but in the WiS regime, there are too few free waters available. Instead, those donors attach to nearby $\mathrm{Cl}^-$, creating solvent-separated $\mathrm{Li}^+$~-~$\mathrm{Cl}^-$ pairs. Thus every $\mathrm{Li}^+$ adsorbed at the graphite interface drags in accompanying $\mathrm{Cl}^-$ anions through its hydration shell, driving the interfacial Li:Cl ratio toward 1:1. In more dilute electrolytes the excess solvent completes the hydrogen-bond network with additional water molecules and $\mathrm{Li}^+$ adsorption remains dominant.

Figure~\ref{fig:CN} (c--f) shows $N_{\mathrm{hyd}}$ and $N_{\mathrm{SSIP}}$ for the charged graphite systems. At the positively charged electrode ($\sigma^{+}$) the $\mathrm{Li^{+}}$ keeps its almost perfectly tetra-hydrated solvation shell in both the $1$ and $10~\mathrm{mol~kg^{-1}}$ electrolytes, $N_{\mathrm{hyd}}\simeq4$ at all $z$ positions. Electrostatic repulsion keeps $\mathrm{Li^{+}}$ out of the IHP, unlike in the $\sigma^{0}$ and $\sigma^{-}$ systems. However, in the $20~\mathrm{mol~kg^{-1}}$ electrolyte, a small number of $\mathrm{Li^{+}}$ ions penetrate the IHP, causing their hydration number to decrease to $N_{\mathrm{hyd}}\approx3$. The SSIP profile at $\sigma^{+}$ also matches that seen at $\sigma^{0}$, but with a smaller interfacial decrease because fewer $\mathrm{Li^{+}}$ ions are present near the surface.

$\mathrm{Cl^{-}}$ is somewhat insensitive to surface polarity. Its hydration number remains $N_{\mathrm{hyd}}\approx5\!-\!6$ across the entire double layer for 1 and 10 mol~kg$^{-1}$, and decreases slightly to $\approx4.8$ at 20 mol~kg$^{-1}$. SSIPs are virtually absent at 1~mol~kg$^{-1}$, rise to $\sim3$ per anion at 10~mol~kg$^{-1}$, and reach $\sim6$ at 20~mol~kg$^{-1}$. The $N_{\mathrm{hyd}}$ and $N_{\mathrm{SSIP}}$ profiles for $\mathrm{Cl^{-}}$ at the negative electrode begin slightly farther from the graphite, reflecting the marginal electrostatic repulsion. Consistent with the bulk-normalised density profiles, the liquid layering observed at the neutral interface persists upon charging, and the virtually identical ion-coordination motifs at neutral and polarised electrodes indicate that $\mathrm{Li^{+}}$-$\mathrm{Cl^{-}}$ co-adsorption is an intrinsic feature of these systems. 

\begin{figure}[htbp]
  \centering
  \includegraphics[width=\textwidth]{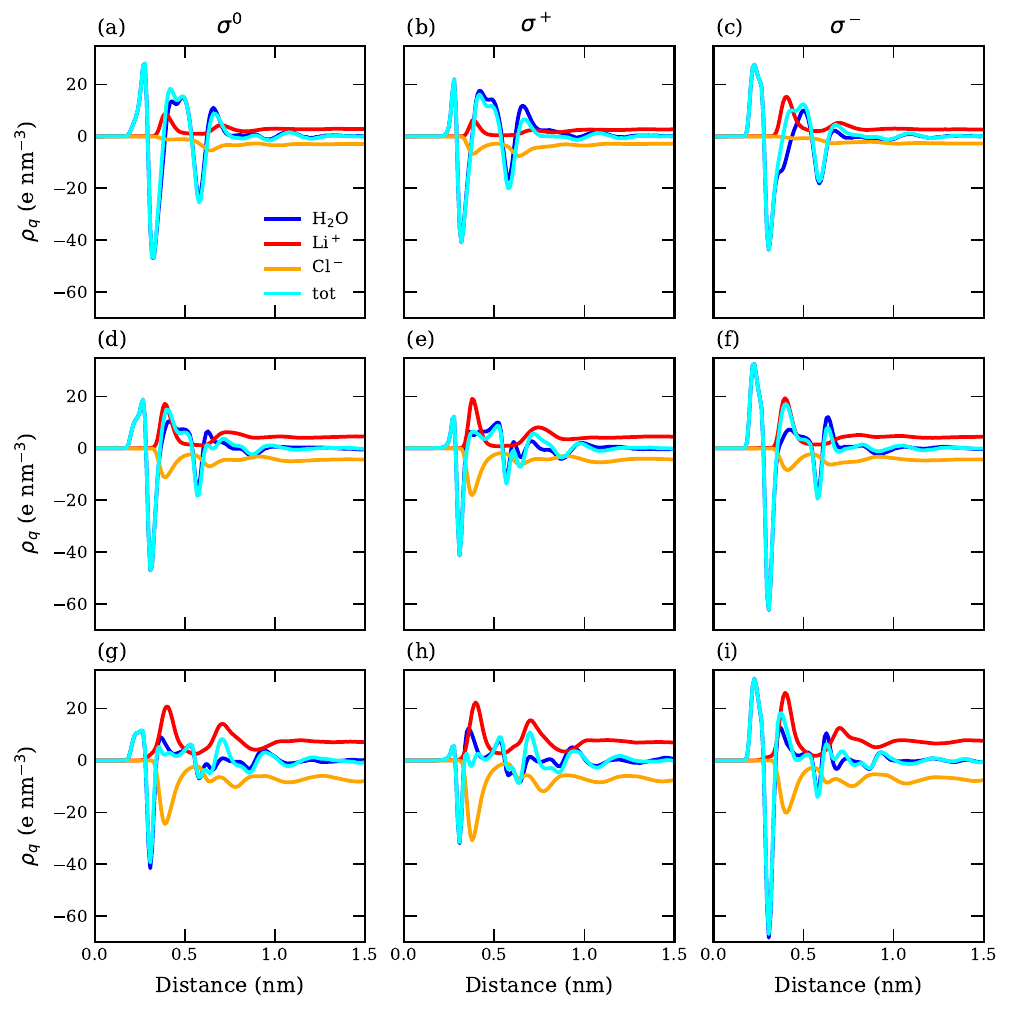}
  \caption{Charge density profiles, $\rho_q^{\alpha}$ (units $e~\mathrm{nm^{-3}}$), for species $\alpha\in\{\mathrm{H_2O},~\mathrm{Li}^+,~\mathrm{Cl}^-\}$ and the total ($\alpha=\mathrm{tot}$), as a function of $z$ distance from the electrode. Columns correspond to electrode surface charge densities $\sigma^0=0~\mathrm{C~m^{-2}}$ (left), $\sigma^+=+0.0571~\mathrm{C~m^{-2}}$ (middle), and $\sigma^-=-0.0571~\mathrm{C~m^{-2}}$ (right). Rows show concentrations: (a--c) $1~\mathrm{mol~kg^{-1}}$ LiCl; (d--f) $10~\mathrm{mol~kg^{-1}}$; (g--i) $20~\mathrm{mol~kg^{-1}}$. Colours: water (blue), $\mathrm{Li}^+$ (red), $\mathrm{Cl}^-$ (yellow), total (cyan).}

  \label{fig:rhoq}
\end{figure}

The counterintuitive features of $\mathrm{Li^{+}}$ adsorption at charged electrodes can be rationalised by considering both the ionic intermolecular bonding environments and the $z$-resolved charge density, $\rho_{\mathrm{q}}(z)$, composed of water and ion contributions.\cite{Dočkal_mixed_alkali} As shown in Figure~\ref{fig:rhoq}, the total charge density, $\rho_{\mathrm{q}}^{\mathrm{tot}}(z)$, closely follows the water contribution, $\rho_{\mathrm{q}}^{\mathrm{H_2O}}(z)$  in all systems, with minor deviations near the OHP. Thus, the interfacial electric field is governed by the spatial and orientational structure of water. 
 
Naturally, the interface suppresses large local charge heterogeneities, constraining how number densities change upon charging. The equilibrium adsorption structure is the free-energy minimum that balances direct graphene-ion/graphene-water interactions with stabilisation from hydration, ion pairing, and water-water interactions, all while suppressing local charge accumulation  encoded in $\rho_{\rm q}^{\rm tot}(z)$.

Because $\rho_{\rm q}^{\rm tot}(z)$ is set primarily by the water structure, the positions of the adsorption layers are largely fixed by solvation and packing. Charging the electrode primarily changes the occupancy of ion adsorption sites, not their location. Across all concentrations and electrode polarities, the first ion adsorption peaks are located in a region of net positive total charge, $\rho_{\mathrm q}^{\mathrm{tot}}(z)>0$. In this framework, Dočkal and co-workers defined an ion as compatible when adsorption at its peak position reduces $\lvert \rho_{\mathrm q}^{\mathrm{tot}}(z)\rvert$, and incompatible when it increases it.\cite{Dočkal_mixed_alkali, Dočkal_understanding} Accordingly, $\mathrm{Cl}^-$ is compatible and its adsorption is electrostatically favourable, as it reduces the local positive charge, whereas $\mathrm{Li}^+$ is incompatible and electrostatically unfavourable, since it increases it. Additionally, the intermolecular bonds between water and $\mathrm{Cl}^-$ are much weaker than those with $\mathrm{Li}^+$, allowing graphite’s electric charge to influence $\mathrm{Cl}^-$ more strongly than $\mathrm{Li}^+$.

The observed trends follow from this picture. At $1~\mathrm{mol~kg^{-1}}$ (Figure~\ref{fig:dens_profile} (a--c); Figure~\ref{fig:rhoq} (a--c)), positive charging markedly increases compatible $\mathrm{Cl}^-$ in the first adsorption layer, while incompatible $\mathrm{Li}^+$ changes negligibly. The total charge density in the OHP--dominated by $\rho_{\mathrm q}^{\mathrm{H_2O}}(z)$--also changes little, so there is no additional local electrostatic penalty for $\mathrm{Li}^+$ beyond its direct repulsion from the electrode. $\mathrm{Cl}^-$ adsorption at the neutral surface is already negligible, so upon negative charging, neutralisation is achieved mainly by water reorientation and enhanced $\mathrm{Li}^+$ adsorption. At $10~\mathrm{mol~kg^{-1}}$ (Figure~\ref{fig:dens_profile} (d--f); Figure~\ref{fig:rhoq} (d--f)), both ions are abundant in the OHP of the neutral surface. Upon positive charging, water reorients so that $\rho_{\mathrm q}^{\mathrm{H_2O}}(z)$--and consequently $\rho_{\mathrm q}^{\mathrm{tot}}(z)$--in the OHP becomes less positive than at neutrality. This lowers the local electrostatic penalty and, counterintuitively, increases $\mathrm{Li}^+$ accumulation in the OHP. In parallel, $\mathrm{Cl}^-$ adsorption in the OHP increases and SSIPs become more prevalent. Under negative charging, $\mathrm{Cl}^-$ adsorption decreases and $\mathrm{Li}^+$ rises only marginally, reflecting the fact that $\rho_{\mathrm q}^{\mathrm{H_2O}}(z)$ in the OHP remains positive and comparable in magnitude to the neutral case, so the electrostatic driving force changes little. In the WiS $20~\mathrm{mol~kg^{-1}}$ system (Figure~\ref{fig:dens_profile} (g--i); Figure~\ref{fig:rhoq} (g--i)), charging again modulates $\mathrm{Cl}^-$ more strongly than $\mathrm{Li}^+$. Positive charging decreases $\rho_{\mathrm q}^{\mathrm{H_2O}}(z)$ in the OHP and yields a slight increase of $\mathrm{Li}^+$ together with a stronger increase of $\mathrm{Cl}^-$. Unlike at $10~\mathrm{mol~kg^{-1}}$, the negative electrode produces a more intuitive increase of $\mathrm{Li}^+$, consistent with the smaller amplitude of $\rho_{\mathrm q}^{\mathrm{H_2O}}(z)$ in the OHP, which reduces the role of compatibility.

\subsection{Capacitance}
\begin{figure}[htbp]
    \centering       \includegraphics{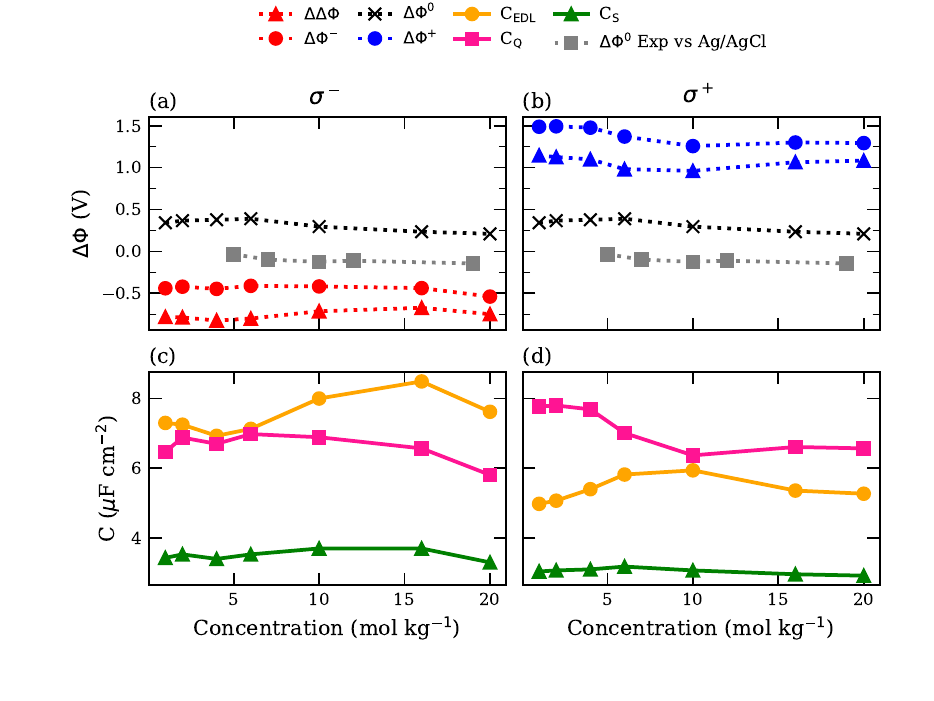}
    \caption{Concentration dependence of the interfacial potential drop and the surface-capacitance components of $\mathrm{LiCl}$-graphite capacitor. (a)~and~(b) show the electrostatic potential difference $\Delta\Phi$ between the electrode plane and the bulk electrolyte for a negatively charged surface ($\sigma^- = -0.057~\mathrm{C~m^{-2}}$, (a) and a positively charged surface ($\sigma^+ = +0.057~\mathrm{C~m^{-2}}$, (b). Black crosses: $\Delta\Phi^{0}$ at the potential of zero charge~(PZC); coloured circles: $\Delta\Phi^{\pm}$ at the charged electrodes; coloured triangles: the excess drop $\Delta\Delta\Phi = \Delta\Phi^{\pm}-\Delta\Phi^{0}$; grey squares: $\Delta\Phi^{0}$ measured from experimental differential capacitance vs an Ag/AgCl reference. (c)~and~(d) give the integral capacitances: the surface capacitance $C_{\mathrm{S}}$ (green), the integral double-layer capacitance $C_{\mathrm{EDL}}$ (pink), and the quantum capacitance $C_{\mathrm{Q}}$ (orange) for $\sigma^-$ (c) and $\sigma^+$ (d). Dotted lines are guides to the eye. Error bars correspond to one standard error from block-averaging but are smaller than the symbol size and have therefore been omitted for clarity.}
    \label{fig:capacitance}
\end{figure}

%pzc (reference to Rob's work)
The thickness of the EDL and its structure have a direct effect on the capacitance value (C$_{\mathrm{S}}$), which can be calculated as the sum of the EDL (C$_{\mathrm{EDL}}$) and quantum ($C_{\mathrm{Q}}$) capacitance,
\begin{equation}
  \frac{1}{C_{\mathrm S}} = \frac{1}{C_{\mathrm Q}} + \frac{1}{C_{\mathrm{EDL}}}.
  \label{cap_series}
\end{equation}
Both terms were extracted from QMMD simulations. The (integral) double-layer capacitance, $C_{\mathrm{EDL}}$, which quantifies charge accumulation in the EDL, is obtained from
\begin{equation}
  C_{\mathrm{EDL}} = \frac{\sigma}{\Delta\!\Delta\Phi},
\end{equation}
where $\sigma$ is the surface charge density and $\Delta\!\Delta\Phi = \Delta\Phi - \Delta\Phi^{0}$ is the change in the interfacial electrostatic potential drop upon charging. The value of $\Delta\Phi$ is computed as $\Delta\Phi = \Phi^{\text{electrode}}-\Phi^{\text{bulk}}$, where $\Phi^{\text{electrode}}$ is the electrostatic potential at the electrode surface and $\Phi^{\text{bulk}}$ is its value in the liquid bulk. The reference value $\Delta\Phi^{0}$ corresponds to $\Delta\Phi$ for a neutral electrode, otherwise known as the PZC. The potential profile $\Phi(z)$ is calculated from the planar-averaged charge density of all liquid species $\rho_{\mathrm q}(z)$ by integrating Poisson’s equation in one dimension,
\begin{equation}
  \Phi(z) = -\frac{1}{\varepsilon_{0}} \int_{0}^{z}\!\! \mathrm dz'~(z-z')~\rho_{\mathrm{q}}(z').
\end{equation}
The differential quantum capacitance is obtained from the electronic density of states, $D(E)$, via
\begin{equation}
  C_{\mathrm Q}^{\mathrm{diff}}(\Phi)
  = \frac{e^{2}}{4 k_{\mathrm B} T}
    \int_{-\infty}^{\infty}
      D(E)~
      \operatorname{sech}^{2}\!\left[\frac{E+\Phi}{2 k_{\mathrm B} T}\right]
      \mathrm dE,
  \label{eq:CQdiff}
\end{equation}
where $E$ is measured relative to the Fermi energy, $E_{\mathrm F}$, $e$ is the elementary charge, $k_{\mathrm B}$ the Boltzmann constant, $T$ the temperature, and $\Phi$ the electrostatic potential of the electrode. $D(E)$ was calculated as a trajectory average from each SCC-DFTB iteration using a $\Gamma$-centred Monkhorst-Pack $8\times8\times1$ k-point grid. The integral (charge-averaged) quantum capacitance that enters the series formula can be estimated as
\begin{equation}
  C_{\mathrm Q}
  = \frac{1}{\Delta\Phi}\int_{0}^{\Delta\Phi}
      C_{\mathrm Q}^{\mathrm{diff}}\!\left(\Phi\right)~
      \mathrm d\Phi.
  \label{eq:CQint}
\end{equation} 
Plots of $\Delta\Phi^{0}$, $\Delta\Phi^{+}$, and $\Delta\Phi^{-}$ are shown in Figure~\ref{fig:capacitance} for the negative ((a)) and positive ((b)) electrodes, together with the corresponding experimental $\Delta\Phi^{0}$ values, obtained from the minima of fourth-order polynomial fits to the measured differential capacitance curves against Ag/AgCl (3.5 M KCl). $C_{\mathrm{EDL}}$, $C_{\mathrm{Q}}$ and $C_{\mathrm{S}}$ for the negative and positive systems are presented in Figures~\ref{fig:capacitance} (c) and (d), respectively. The PZC, $\Delta\Phi^{0}$, increases up to $6~\mathrm{mol~kg^{-1}}$ as the $\mathrm{Li}^+$ peak in the OHP grows while the small $\mathrm{Cl}^-$ peak remains nearly unchanged, thereby increasing the net positive charge stored in the double layer. Beyond $6~\mathrm{mol~kg^{-1}}$, however, accumulation of $\mathrm{Cl}^-$ in the OHP offsets the positive charge from adsorbed $\mathrm{Li}^+$, leading to a decrease in $\Delta\Phi^{0}$ (Figure~\ref{fig:capacitance} (a, b)). Experimentally, the PZC decreases gradually between $5$ and $19 ~\mathrm{mol~kg^{-1}}$, with a total change of $-0.11~\mathrm{V}$. This behaviour is qualitatively consistent with the simulations, which exhibit a decrease of $-0.18~\mathrm{V}$ between $6$ and $20~\mathrm{mol~kg^{-1}}$. Finney \textit{et al.} likewise reported a concentration-dependent decrease in the PZC for aqueous NaCl at graphite, corroborated by simulation and experiment, indicating that this behaviour may also be applicable to other alkali halide salts at graphitic interfaces. Their work further showed that the EDL thickness follows a non-monotonic trend, decreasing before increasing beyond a threshold concentration due to oscillatory screening. \cite{finney2021electrochemistry, finney2024properties} At the negatively charged electrode, $\Delta\Phi^{-}$ remains nearly constant across all concentrations. As a result, the relationship between the excess drop $\Delta\!\Delta\Phi^-$ and concentration is dominated by the trend of $\Delta\Phi^{0}$, decreasing slightly up to $4~\mathrm{mol~kg^{-1}}$, increasing between $4$ and $10~\mathrm{mol~kg^{-1}}$, and then plateauing at higher concentrations. $C_{\mathrm{EDL}}$ follows a similar trend, while $C_{\mathrm{Q}}$ increases between $1-2~\mathrm{mol~kg^{-1}}$, then shows a steady decrease. Consequently, the total surface capacitance $C_{\mathrm{S}}$ stays nearly constant across the full concentration range (Figure~\ref{fig:capacitance}). For the positive electrode, $\Delta\Phi^{+}$ is constant up to $4~\mathrm{mol~kg^{-1}}$, falls between $4$ and $10~\mathrm{mol~kg^{-1}}$, and then plateaus. In combination with the shifting $\Delta\Phi^{0}$, this makes $\Delta\!\Delta\Phi^+$ fall from $1-6~\mathrm{mol~kg^{-1}}$ and rise thereafter; $C_{\mathrm{EDL}}$ shows the inverse behaviour to $\Delta\!\Delta\Phi^+$, while $C_{\mathrm{Q}}$ is essentially constant between $1-4~\mathrm{mol~kg^{-1}}$, decreases between $4-10~\mathrm{mol~kg^{-1}}$ and then plateaus at higher concentrations. Again, the resulting $C_{\mathrm{S}}$ remains essentially independent of concentration (Figure~\ref{fig:capacitance} (d)).

Notably, as additional graphene layers are added, the quantum capacitance $C_{\mathrm{Q}}$ increases significantly.\cite{zhan2015quantum, yin2024unraveling} Therefore, the $C_{\mathrm{EDL}}$ contribution in the series relation (Equation~\ref{cap_series}) becomes more dominant. In the absence of an electronic bottleneck, $C_{\mathrm{S}}$ would simply track $C_{\mathrm{EDL}}$, and would increase in line with the structural changes to the EDL. Figures~\ref{fig:capacitance} (a--b) also reveal that the reference PZC, $\Delta\Phi^{0}$, varies appreciably with concentration and strongly influences the excess drop $\Delta\!\Delta\Phi$. 
For $\sigma^-$, $\Delta\Phi^{-}$ is almost concentration-independent, yet the decrease in $\Delta\Phi^{0}$ beyond $6~\mathrm{mol~kg^{-1}}$ narrows $\Delta\!\Delta\Phi$ and yields the apparent rise in $C_{\mathrm{EDL}}$. This dependence provides the mechanistic insight that changes at the neutral interface modify the excess potential required to reach a given $\sigma$ and thus the integral capacitance. However, when capacitances from different electrolytes are compared, a common potential reference should be considered; otherwise, PZC shifts can be mistaken for genuine changes in performance.

\section{Conclusions}
In this work, molecular dynamics simulations combined with quantum-mechanical molecular dynamics have elucidated the intricate structural and electrochemical behaviour of concentrated and water-in-salt (WiS) aqueous LiCl solutions at graphitic electrodes. Across a broad concentration range ($1-20\;\mathrm{mol~kg^{-1}}$), significant alterations in the electrical double layer (EDL) structure and ion distribution were observed, reflecting a complex interplay among ionic adsorption, hydration, and solvent-separated ion pair (SSIP) formation.

Bulk-normalised density profiles reveal pronounced changes in interfacial structure in concentrated and WiS LiCl solutions. At low concentrations ($<6~\mathrm{mol~kg^{-1}}$), $\mathrm{Li}^+$ preferentially adsorbs at the electrode, resulting in a net positive charge within the outer Helmholtz plane (OHP) and an increase in the potential of zero charge (PZC). As the concentration rises beyond $6~\mathrm{mol~kg^{-1}}$, the number density profiles show substantial $\mathrm{Cl}^-$ accumulation at the interface, driven by the formation of SSIPs and the limited availability of free water, yielding nearly overlapping $\mathrm{Li}^+$ and $\mathrm{Cl}^-$ density peaks in the WiS regime ($16\!-20~\mathrm{mol~kg^{-1}}$) and, with minimal contact ion-pairs.

These structural changes lead to an anomalous trend in EDL thickness as a function of electrolyte concentration. Unlike the classical Debye length prediction, which anticipates a monotonic decrease with increasing ion concentration, the EDL thickness exhibited a non-monotonic and electrode charge-dependent behaviour. At neutral electrodes, the thickness initially decreases but subsequently increases beyond $4~\mathrm{mol~kg^{-1}}$, driven by a lack of available free water and enhanced SSIP formation, leading to ion co-adsorption. Positively charged electrodes exhibit a similar but earlier (lower concentration) transition due to $\mathrm{Cl}^-$ ion accumulation inside the OHP for concentration as low as $2~\mathrm{mol~kg^{-1}}$, whereas negatively charged electrodes show a monotonic increase in thickness with concentration.

These interfacial structural changes directly impact the electrostatic potential drop at the electrode-electrolyte interface and the capacitance values. As the concentration increases, the overlap of $\mathrm{Li}^+$ and $\mathrm{Cl}^-$ density profiles in the OHP leads to enhanced charge compensation at the interface, reducing the net potential drop $\Delta\Phi^{0}$ at the PZC. This is in qualitative agreement with experimental measurements. For negatively charged electrodes, the potential drop at the surface remains nearly constant across concentrations, while shifts in the PZC primarily influence the excess drop $\Delta\!\Delta\Phi$. At positively charged electrodes, both the surface potential drop and $\Delta\!\Delta\Phi$ show a clearer dependence on electrolyte concentration, reflecting the earlier onset of anion accumulation. Despite the pronounced concentration-dependent trends in both EDL and quantum capacitances, their opposing behaviours cause the total surface capacitance, $C_{\mathrm{S}}$, to remain largely insensitive to electrolyte concentration. Despite this, as the quantum capacitance, $C_{\mathrm{Q}}$, increases with the number of graphene layers, the total capacitance approaches the EDL capacitance, $C_{\mathrm{EDL}}$, in multilayer systems. These findings emphasise the critical role of interfacial structure in governing the electrochemical response and highlight the need to specify reference potentials when comparing capacitances across different electrolyte concentrations.

\section{Acknowledgements}
The work was conducted with support from EPSRC Centre for Doctoral Training “Graphene NOWNANO” (grant EP/L01548X/1). R.A.W.D. would like to thank the EPSRC for further support (grants EP/T01816X/1 and EP/V049925/1). Simulations were performed on the Computational Shared Facility at the University of Manchester.

\clearpage

\bibliography{refs}
\end{document}